\documentclass[aip,graphicx]{revtex4-1}

\usepackage{graphics}
\usepackage{amsmath}
\usepackage{xcolor}
\usepackage{soul}

\newcommand\redsout{\bgroup\markoverwith{\textcolor{red}{\rule[0.5ex]{2pt}{0.4pt}}}\ULon}

\draft 

\begin{document}



\title{Towards Flexible Intensity Control of Resonantly Scattered $\gamma$-Rays Using Multi-Frequency Vibrating Resonant Absorber}



\author{Ale\v{s} Stejskal}
\author{Vlastimil Vrba}
\author{V\'{i}t Proch\'{a}zka}
\email[]{v.prochazka@upol.cz}
\affiliation{Department of Experimental Physics, Faculty of Science, Palack\'{y} University Olomouc, 17. listopadu 1192/12, 779 00 Olomouc, Czech Republic}


\date{\today}

\begin{abstract}
	We report a method for coherent control of $\gamma$-photons, enabling the shaping of $\gamma$-ray intensity in nearly arbitrary waveforms. Different intensity waveforms are created by adjusting the motion profile of a resonant absorber (an ensemble of M\"{o}ssbauer nuclei) and tuning the energy of the incident radiation. A crucial aspect of this method is the use of a low fundamental frequency of vibrations, which broadens the possibilities for $\gamma$-ray control. The results of numerical simulations are experimentally validated by generating single and double $\gamma$-pulses and inducing short-term absorption. For this, a resonant absorber containing  $^{57}$Fe nuclei was vibrated with different motion profiles composed of 12 harmonics with a fundamental frequency of 1\,MHz. The proposed technique represents an advancement in the manipulation of $\gamma$-rays and potentially X-rays, paving the way for the performance of unique types of $\gamma$-ray or X-ray quantum experiments and the development of tools such as adjustable tabletop $\gamma$-pulse sources or $\gamma$-ray or X-ray delays and gates. Moreover, inverse application of the method enables investigation of motion at the picometer scale.  
\end{abstract}


\maketitle 


Coherent control of $\gamma$-ray and hard X-ray photons (10--100\,keV) is of interest for their unique properties, including high frequency, short wavelength, deep material penetration, and high-efficiency and low-noise detection. For that reason, the potential applications of high-energy radiation are explored in nuclear quantum optics \cite{Kuznetsova2017,Heeg_2017,Bernhard2012}, quantum technologies \cite{Kocharovskaya2019_memory,Shakhmuratov2015_pulses,Liao2012,Pallfy2009}, and metrology  \cite{Shakhmuratov2020_granular_vibrations,Goerttler2019_metrology, Shakhmuratov2018, Wense2018_nuclear_clock}. However, nuclear quantum experiments contend with a lack of coherent radiation sources. Therefore, different concepts of coherent control of $\gamma$-photons have been developed, such as electromagnetically or acoustically induced transparency \cite{Rohlsberger2012,Radeonychev2020_transparency,Coussement2002_EIT_anticrossing}. These concepts are based on influencing the recoilless resonant interaction of the radiation with nuclear transition using electromagnetic fields or ultrasound vibrations. Particularly, acoustical manipulation has potential in long-lifetime nuclear systems, making it suitable for the control of $\gamma$-photons. Desired interference effects occur when a resonant absorber vibrates at frequencies comparable to the excited state lifetime, and the amplitude of vibrations is comparable to the $\gamma$-photon wavelength.

Several effects of ultrasound vibrations on $\gamma$-rays were described in the time and energy domains. For example, super-radiant states, also called gamma echoes, can be generated by step-like absorber displacements \cite{Shakhmuratov2014_superradiance,Shakhmuratov2011,Helisto1991_echo}. Next, a decomposition of the single-line absorber spectrum into a comb of several equidistant spectral components (sidebands) can be induced by sinusoidal vibrations \cite{Shakhmuratov2017_combs,Ruby1960}. In the time domain, the same motion generates periodic $\gamma$-pulses arising from the pulse-shaped structures in the single photon wave packet \cite{Vagizov2014}. This approach also allows the production of bunches of $\gamma$-pulses \cite{Shakhmuratov2015_pulses}. However, sinusoidal motion provides only limited control of the temporal properties (intensity, duration) of the $\gamma$-pulses.  

Within the last six years, two theoretical works were published on enhancing the temporal properties, namely the intensity and width, of the $\gamma$-pulses in the single photon wave packet \cite{Khairulin2023,Khairulin2018}. In these papers, adding more harmonic frequencies to the absorber motion profile was proposed, and the optimal motion to generate the most intense pulses was derived. However, the implementation of these experiments faces challenges due to the requirement of high fundamental frequencies ($\geq$30\,MHz) of the vibrational motion.

In this Letter, we present a generalized concept of $\gamma$-ray intensity control by a vibrating absorber whose motion profile is composed of several harmonic frequencies. This approach allows the constant intensity of $\gamma$-rays from the radioactive source to be shaped into a wide variety of waveforms by adjusting the motion profile of the absorber and the energy of the incident $\gamma$-photons. Thus, it allows the generation not only of single pulses but also multi-pulse structures or short-term absorption of different durations and intensity levels. The utilization of low fundamental frequencies is crucial for this purpose, as it enables long repetition periods ($>$100\,ns), large off-duty cycles ($>$99\,\%, part of the period with intensity suppressed below half maximum of the pulse), and higher pulse intensities. The low frequencies, however, present a challenge in the analytical description due to the overlap of spectral sidebands \cite{Khairulin2023}. These complications are addressed with numerical calculations, which we validated experimentally. We anticipate that this technique will enable unique types of $\gamma$-optics experiments, e.g., of the pump-probe\cite{Heeg2021} type, or the development of tools, such as adjustable table-top $\gamma$-pulse sources or $\gamma$-ray delays and gates. The principles of this technique are also applicable to manipulation with hard X-rays. Moreover, the method can be inversely used for the investigation of motion at the picometer level.

The coherent control of $\gamma$-photons emitted from the radioactive source by the ultrasound vibrations of the absorber can be described using a semi-classical approach \cite{Vagizov2014}.  Thus, in the laboratory reference frame, the $\gamma$-photon field incident on the absorber is represented as a classical quasi-monochromatic wave
\begin{equation} \label{Eq1}
\begin{split}
E(z,t) = E_{0}\theta\left(t-t_{0}-\frac{z}{c}\right) \hspace{3.5cm}\\
\times\exp{\left[-\left(i\omega_{r}+\frac{\Gamma}{2}\right)\left(t-t_{0}-\frac{z}{c}\right) + i\varphi_{0}\right]},
\end{split}
\end{equation}
where $E_{0}$ is the wave amplitude, $\theta$ is the Heaviside step function, $\omega_{r}$ is the resonant frequency, $i$ is the imaginary unit, $\Gamma$ is the inverse of the excited state lifetime, $t_{0}$ is the origin time of the excited state, $c$ is the speed of light in vacuum, $\varphi_{0}$ is an initial random phase, and $z$ is the absorber coordinate along the $\gamma$-photon propagation direction. In our case, we consider general periodic and uniform absorber motion $z(t)$ which can be expressed by the finite Fourier series
\begin{equation} \label{Eq2}
z(t) = z_{0} + \frac{pc}{\omega_{r}}\sum_{k=1}^{K}a_{k}\sin\left(2\pi k f t + \phi_{k}\right),
\end{equation}
where $z_{0}$ is the rest distance between the absorber and the radioactive source, $a_{k}$ and $\phi_{k}$ are the normalized Fourier coefficients of the motion profile ($\max|\sum_{k=1}^{K}a_{k}\sin\left(2\pi k f t+\phi_{k}\right)| =1$),  $K$ specifies the number of harmonics used to realize the motion ($a_{k} = 0$ for $k>K$), $f$ is the fundamental frequency of periodic motion, and $p$ is a dimensionless parameter (amplitude modulation index \cite{Vagizov2014}) proportional to the vibrations amplitude $A$ by relation $p = 2\pi A/ \lambda$,  where $\lambda$ is the $\gamma$-photon wavelength.

Equations~(\ref{Eq1}) and~(\ref{Eq2}) describe the $\gamma$-photon field in the reference frame of the vibrating absorber. Based on these equations, the integral normalized radiation intensity waveform $I/I_{0}$ of the resonant $\gamma$-photons detected behind the absorber can be calculated numerically via Fourier transform, and integration over the random time of the excited state ($\gamma$-photon) origin~$t_{0}$. $I_{0}$ denotes the incident $\gamma$-ray intensity of the resonant photons. For more details, see the Supplemental Material, where the numerical calculations are described in detail and the corresponding Python script is attached.   

The shape of radiation intensity waveform depends on the following parameters: absorber motion profile $z(t)$, energy detuning of the incident radiation from the resonance $\Delta$ (energy within this article is expressed in megahertz units, where 1\,MHz = 0.086\,mm/s = 4.14\,neV), absorber hyperfine parameters, and optical thickness $T_{a}$. In the context of simulations, our objective was the examination of the motion profiles using low fundamental frequencies ($f < 5$\,MHz) and different energy detuning. An unlimited number of different motion profiles provide innumerable possibilities for manipulation with $\gamma$-rays. Therefore we chose only four example shapes: the square, trapezoid, saw-tooth, and bipolar pulse (a pulse in one direction followed by a pulse in the opposite direction). For each profile, the amplitude, number of harmonics, and fundamental frequency were varied. The $^{57}$Fe M\"{o}ssbauer isotope was considered as the absorber, which exhibits an excited state lifetime of 1/$\Gamma \approx 141$\,ns and a transition energy of 14.41\,keV.
\begin{figure*}[]
	\includegraphics{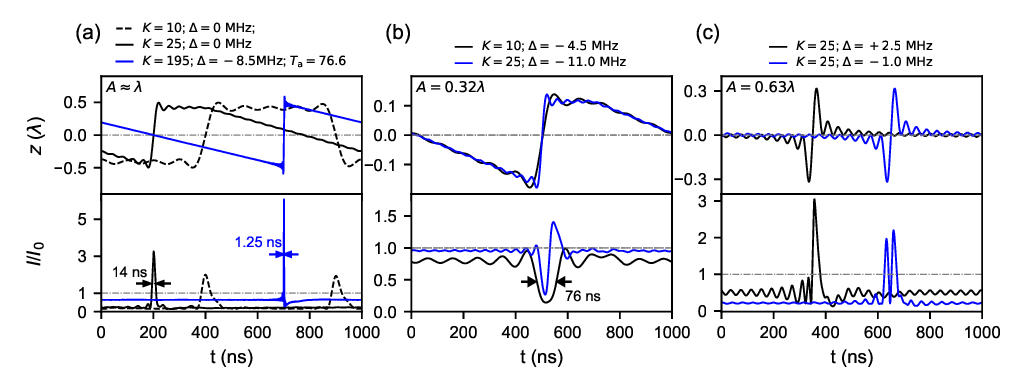}
	\caption{Simulations of normalized intensity of the resonant $\gamma$-photons behind the absorber $I/I_{0}$ for different absorber motion profiles $z(t)$, number of harmonics $K$, and energy detuning $\Delta$. (a) Generation of single $\gamma$-pulses using square (black-dashed line), trapezoidal (black solid line) and saw-tooth motion profile (blue solid line). (b) Short-time induced absorption using trapezoidal profile. (c) Generation either a single pulse or double pulse by changing the energy detuning using bipolar pulse profile. Simulations consider $T_{a}=14$ (if otherwise noted) corresponding to approximately 1\,$\mu$m thick $^{57}$Fe absorber. Absorber displacement $z$ is expressed in units of $\gamma$-photon wavelength $\lambda$ equal to 86\,pm.
		\label{Fig1}}
\end{figure*}

The single $\gamma$-pulses are created if the absorber motion profile contains rapid displacements and the incident radiation energy is near resonance ($\Delta \approx 0$\,MHz). Examples of such motion profiles: the square, trapezoidal, and saw-tooth are shown in Fig.~\ref{Fig1}(a).
Pulse width is proportional to displacement duration, so the minimum width is restricted by highest harmonic frequency in the absorber motion profile. The pulse intensity can be higher as the displacement is faster. The highest intensities are obtained with amplitude displacement close to the $\gamma$-photon wavelength ($A = \lambda$). Higher amplitudes do not result in a significant increase in intensity but instead split it into multi-pulse structures. Assuming the same displacement amplitude and number of harmonics, the highest pulses (about 5\,\% higher) are generated by a saw-tooth profile. Similar results were derived in \cite{Khairulin2023}, nevertheless, we also observed them for low fundamental frequencies.

We also searched for optimal conditions (number of harmonics, energy detuning, and optical thickness) for generating the most intense single pulses using a saw-tooth motion profile with up to 250~harmonics and the fundamental frequency 1\,MHz. We compared the obtained results with the conditions from \cite{Khairulin2023}, where the fundamental frequency 30\,MHz and 30 harmonics were suggested. The conditions and pulse characteristics for both cases are summarized in Table~\ref{Table1}. For $f=1$\,MHz, the maximum found intensity was $I = 6.1I_{0}$ [blue curve in Fig.~\ref{Fig1}(a)], while for  $f=30$\,MHz it was $I = 3.8I_{0}$. It is noteworthy that a lower fundamental frequency yielded superior results, even though the frequency spectrum of the absorber motion profile was 4.6 times narrower (195\,MHz instead of 900\,MHz). This makes it easier to realize such motion experimentally.

\begin{table*}[]
	\caption{\label{Table1} Comparison of optimal parameters for generating the most intense pulses between this work and Ref. \cite{Khairulin2023}.  
	}
	\begin{ruledtabular}
		\begin{tabular}{l c c c c c c c c }
			& $f$ (MHz) & $K$  & $\Delta$ (MHz) & $T_{a}$ && $I/I_{0}$ & Pulse width\footnote{FWHM - full width at half maximum} (ns) & Off-duty cycle\footnote{with respect to pulse FWHM} (\%) \\
			\hline
			Ref. \cite{Khairulin2023} & 30 & 30 & $-$30 & 45.7 && 3.8 & 0.39 & 98.8 \\
			This work & 1 & 195 & $-$8.5 & 76.6 && 6.1 & 1.25 & 99.9 \\
		\end{tabular}
	\end{ruledtabular}
\end{table*}

In addition to the single $\gamma$-pulses, the rapid displacements of lower amplitudes allow inducing a short-term absorption ($<$100\,ns) when the incident energy is tuned out of resonance, see Fig.~\ref{Fig1}(b). The velocity of the absorber during displacement causes Doppler energy modulation, temporarily tuning the incident radiation energy to resonance. Due to the complexity of the interference process, the absorption is followed by an intensity overshoot which is more significant with shorter displacement durations, see the blue curve in Fig.~\ref{Fig1}(b).

The bipolar pulse motion profile is noteworthy as it creates a double pulse instead of a single pulse only by changing the energy detuning by 3.5\,MHz (0.3\,mm/s) as seen in Fig.~\ref{Fig1}(c). Such double pulses could be of interest in the context of single-photon entanglement as time qubits \cite{Pallfy2009}.

\begin{figure}
	\includegraphics{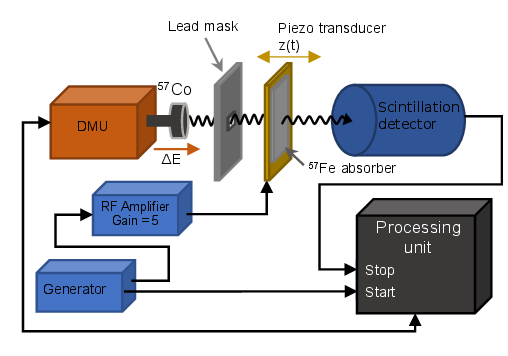}
	\caption{Experimental setup for coherent control of $\gamma$-ray intensity. The $\gamma$-rays from the radioactive source $^{57}$Co, collimated by a lead mask, are scattered by the vibrating $^{57}$Fe-containing absorber, and then registered by a scintillation detector. The $\gamma$-ray intensity is measured relative to the start of the absorber motion period using a processing unit (M\"{o}ssbauer spectrometer) with an integrated time-to-digital converter \protect\cite{Stejskal2023}. Energy of the incident radiation is tuned by a Doppler modulation unit (DMU) in constant velocity mode.
		\label{Fig2}}
\end{figure}

Simulations were experimentally validated using the M\"{o}ssbauer transmission setup shown in Fig.~\ref{Fig2}. The low-amplitude, high-frequency motion of the absorber was realized by polyvinylidene fluoride (PVDF) piezo transducer driven by an amplified voltage waveform from an arbitrary function generator. The absorber was stainless steel foil (type 304) with natural abundance of $^{57}$Fe (2.119\,\%). Analysis of the absorber by means of transmission M\"{o}ssbauer spectroscopy with resonant detector  \cite{Prochazka2022} provided us with information about absorber weak hyperfine magnetic field $0.58 \pm 0.02$\,T and optical (effective) thickness $T_{a}= 7.6 \pm 0.3$. Both values were necessary for data fitting. For more technical details regarding the experimental setup, see Supplemental Material. 

To achieve the particular motion profile of the absorber, it is essential to measure the amplitude and the phase of the absorber motion as a function of frequency and driving voltage (frequency response function). The easiest way of measuring the picometer-scale absorber displacements is fitting the $\gamma$-ray intensity waveforms influenced by the motion itself \cite{Goerttler2019_metrology}. To do this with a radioactive source, we used sinusoidal motion of the absorber and measured the $\gamma$-ray intensity in the time domain \cite{Vagizov2014}. By fitting the intensity waveforms for specific frequencies and amplitudes of the driving voltage, the amplitudes and phases of the harmonic absorber motion were obtained. Thus, the response function was measured point by point for 12~frequencies from 1\,MHz to 12\,MHz. Higher frequencies were excluded due to relatively low amplitude response. More details regarding the frequency response function are provided in the Supplemental Material.

Assuming the linearity of the piezo transducer (experimentally validated to a reasonable degree), we calculated the voltage waveforms $V_{\mathrm{d}}(t)$ from the frequency response function to realize desired absorber motion profiles $z(t)$ assembled from 12~harmonics. The amplitudes and phases of these harmonics were obtained from the Fourier decomposition of the required ideal smooth motion profile (trapezoidal, square) normalized to unit amplitude. The higher amplitude motion profile was then obtained by multiplying the amplitudes of the individual harmonics by a constant.  

\begin{figure}[]
	\includegraphics{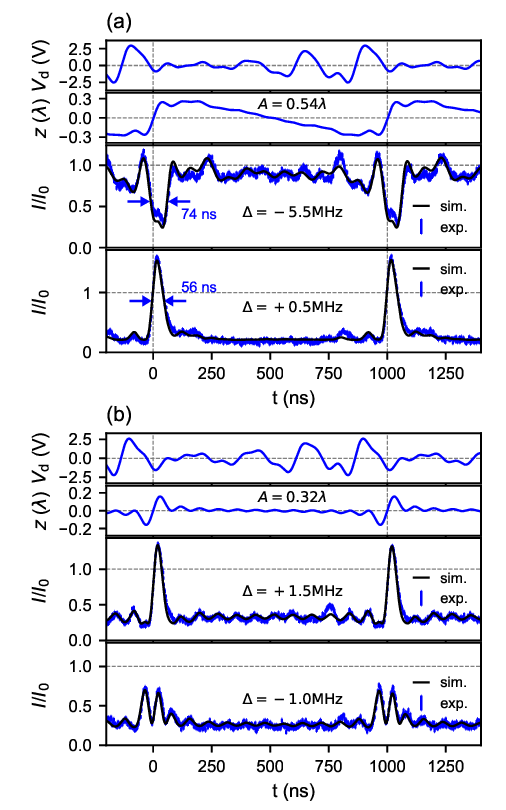}
	\caption{Experimental results of $\gamma$-ray intensity control compared to simulations. (a) Induction of absorption for 74\,ns at $\Delta = -5.5$\,MHz and generation of 56\,ns wide single $\gamma$-pulses at $\Delta = +0.5$\,MHz  by trapezoidal motion profile. (b) Generation of single and double $\gamma$-pulses by bipolar pulse motion profile at $\Delta = +1.5$\,MHz and $\Delta = -1.0$\,MHz, respectively. $V_{d} (t)$ denotes voltage waveform set on generator which is expected to realise the absorber motion profile $z(t)$.
		\label{Fig3}}
\end{figure}

Application of the voltage waveform for trapezoidal motion profile induced absorption with the minimum level 0.3$I_{0}$ for 76\,ns when the energy was detuned off the resonance by $-$5.5\,MHz, and generated 56\,ns wide (full width at half maximum)  single $\gamma$-pulses with intensity 1.7$I_{0}$ when incident photon energy was tuned near the resonance ($\Delta = 0.5$\,MHz), see Fig.~\ref{Fig3}(a). The achieved off-duty cycle was 7.4\,\%  and 94.4\,\% for absorption and $\gamma$-pulses, respectively. Generation of either double or single $\gamma$-pulse in dependence on the energy detuning is shown in Fig.~\ref{Fig3}(b). Presented experimental data are normalized to incident intensity of resonant $\gamma$-photons to be directly compared with simulations, see the Supplemental Material for more details.  High degree of agreement between experiments and simulations in Fig.~\ref{Fig3} confirms the validity of numerical calculations. Occasional discrepancies are caused by deviations of the real absorber motion from the expected motion profile.

\begin{figure}[]
	\includegraphics{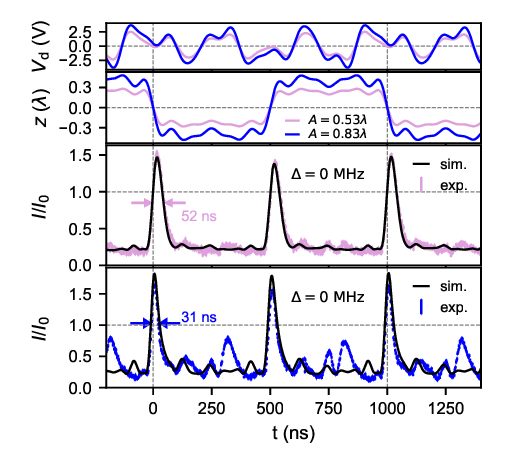}
	\caption{Experimental results of $\gamma$-ray intensity control by a square motion profile. Discrepancy between the experiment and the simulation increases as absorber motion amplitude rises from $A = 0.53 \lambda$ to $A = 0.83 \lambda$, see the violet and the blue curves, respectively.
		\label{Fig4}}
\end{figure}

We aimed to enhance the $\gamma$-pulse intensity by increasing rapid absorber displacement towards the optimal value $A = \lambda$. Fig.~\ref{Fig4} illustrates these attempts with a square motion profile. Higher displacement reduced the pulse width to 31\,ns but discrepancy between the experiment and simulation increased (blue curves in Fig.~\ref{Fig4}). This suggests that the behaviour of the absorber differs from the predicted pattern. We hypothesize that the absorber motion might be too non-uniform \cite{Shakhmuratov2017_combs} at specific frequencies when reaching higher amplitudes. Since we only take into account uniform motion of the absorber, the frequency response function could be incorrectly measured at particular points, giving erroneous driving voltage waveforms. The influence of the non-uniform motion was not studied in detail; however, we observed that the agreement between experiment and simulation improved when the amplitude of the driving voltage was suppressed at frequencies with the minimum amplitude response (specifically, 4\,MHz and 5\,MHz). The uniformity of the motion could be improved by using the lead mask with a smaller aperture, but this undesirably reduces the total flux of the $\gamma$-rays. For that reason, we think that a better way is using a different type of piezo element or modifying the transducer frequency response to ensure uniform motion of the whole absorber. In this context, application of enriched absorbers could be helpful not only due to lower electron absorption but also due to a reduction in the absorber weight, which could improve the transducer frequency response function. 

The pulse shapes could be further improved by finding the optimal motion profile, as we investigated only the four basic motion profiles here. Hypothetically, applying a more complex motion profile could generate more symmetrical pulses with minimal ripple between them. Such a profile could be found by optimization algorithms, which are currently the subject of our further investigation. Similarly, the optimal motion profile could also be determined for rectangular-shaped acoustically induced absorption of defined duration or generally any intensity profile.  

The introduced concept could be further extended by the theoretical idea presented in \cite{Radeonychev2015_abs_sequence}, where more intense and narrower $\gamma$-pulses are proposed using several harmonically vibrating absorbers. In contrast to the mentioned study, using multi-frequency vibrating absorbers could achieve the same results or even enhance them with a lower number of absorbers.

In summary, the $\gamma$-ray intensity waveform can be variously shaped by vibrating the resonant absorber with an appropriate motion profile and tuning the energy of the incident radiation. The utilization of low fundamental frequencies of absorber motion is crucial because it broadens the possibilities of intensity control. Challenges regarding the analytical description can be overcome by numerical calculations. This concept was demonstrated experimentally by generating single and double $\gamma$-pulses and short-term induced absorption using four different absorber motion profiles composed of 12 harmonics with fundamental frequency of 1\,MHz. The experimental achievement of some simulation results like the pulses of higher intensity is currently restricted by the technical capabilities of used piezo transducer, which must ensure required absorber motion. Finally, we suggest a few improvements of this concept such as utilization of enriched absorbers, employment of optimization algorithms for searching optimal motion profiles or utilization of several vibrating absorbers in the row. 

\section*{Supplementary Material}
	The supplementary material contains a description of the numerical calculation of $\gamma$-ray intensity passing through a multi-frequency vibrating absorber. The calculations are also simplified to optimize the computation time by avoiding negligible expressions. Furthermore, details of the experimental setup, mainly the piezo-transducer, are provided. Next, the procedure of measuring the frequency response function of the piezo-transducer is described. Finally, the normalization of experimentally measured $\gamma$-ray intensity is specified so it can be compared with simulations. 

\begin{acknowledgments}
	Authors thank to internal IGA grant of Palack\'{y} University (IGA\_PrF\_2024\_002) and grant no. JG\_2023\_030 of Palack\'{y} University Olomouc for financial support.
\end{acknowledgments}

\section*{AUTHOR DECLARATIONS}
\subsection*{Conflict of Interest}
The authors have no conflicts to disclose.

\section*{Data Availability Statement}
The data that support the findings of this study are available from the corresponding author upon reasonable request.


%
%

%


\bibliography{bibliography.bib}

\end{document}